\documentclass[11pt]{article}
\pdfoutput=1
\newcommand{\Comment}[1]{{}}
\usepackage{amsfonts,amsthm,amsmath,amssymb,slashed}
\usepackage[textwidth = 430 pt, textheight = 630 pt]{geometry}
\usepackage{color}

\Comment{\usepackage{color}
\definecolor{MyDarkBlue}{rgb}{0.15,0.15,0.45}
\usepackage[linktocpage=true]{hyperref}
\hypersetup{
colorlinks=true,
citecolor=MyDarkBlue,
linkcolor=MyDarkBlue,
urlcolor=MyDarkBlue,
pdfauthor={Marcelo R. Barbosa and Horatiu Nastase},
pdftitle={Penrose limit of T-dual of MNa solution and dual orbifold field theory},
pdfsubject={hep-th}
}
\usepackage[utf8]{inputenc}
\usepackage[numbers,sort&compress]{natbib}
\usepackage{hypernat}}
\usepackage{graphicx}
\usepackage{cite}
\usepackage[colorlinks=true]{hyperref}
\hypersetup{
  allcolors = blue,
}

\newcommand\ignore[1]{}
\def\one{{\,\hbox{1\kern-.8mm l}}}

\def\Tr{{\rm Tr\, }}

\def\a{\alpha}\def\b{\beta}

\def\d{\partial}

\def\Tr{\mathop{\rm Tr}\nolimits}

\newcommand{\Cset}{{\,\,{{{^{_{\pmb{\mid}}}}\kern-.45em{\mathrm C}}}}}

\newcommand{\be}{\begin{equation}}
\newcommand{\bea}{\begin{eqnarray}}

\newcommand{\ee}{\end{equation}}
\newcommand{\eea}{\end{eqnarray}}

\begin{document}

\renewcommand{\thefootnote}{\fnsymbol{footnote}}

\makeatletter
\@addtoreset{equation}{section}
\makeatother
\renewcommand{\theequation}{\thesection.\arabic{equation}}

\rightline{}
\rightline{}




\begin{center}
{\LARGE \bf{\sc Penrose limit of T-dual of MNa solution and dual orbifold field theory}}
\end{center}
 \vspace{1truecm}
\thispagestyle{empty} \centerline{
{\large \bf {\sc Marcelo R. Barbosa${}^{a}$}}\footnote{E-mail address: \Comment{\href{mailto:mr.barbosa@unesp.br}}{\tt mr.barbosa@unesp.br}}
{\bf{\sc and}}
{\large \bf {\sc Horatiu Nastase${}^{a}$}}\footnote{E-mail address: \Comment{\href{mailto:horatiu.nastase@unesp.br}}{\tt horatiu.nastase@unesp.br}}

                                                        }

\vspace{.5cm}

\centerline{{\it ${}^a$Instituto de F\'{i}sica Te\'{o}rica, UNESP-Universidade Estadual Paulista}}
\centerline{{\it R. Dr. Bento T. Ferraz 271, Bl. II, Sao Paulo 01140-070, SP, Brazil}}



\vspace{1truecm}

\thispagestyle{empty}

\centerline{\sc Abstract}

\vspace{.4truecm}

\begin{center}
\begin{minipage}[c]{380pt}
{\noindent 
In this paper we study the Penrose limit of the T-dual of the MNa solution and its field theory dual, in order to better 
understand the effect of T-duality in this case. We find a matching of string pp wave oscillators and their masses
to the field theory modes, that are rearranged after T-duality. The effect of T-duality on the long "annulon-type" operators is found,
as a symmetry of the 2+1 dimensional confining theory with spontaneous susy breaking.
}
\end{minipage}
\end{center}

\vspace{.5cm}

\setcounter{page}{0}
\setcounter{tocdepth}{2}

\newpage

\renewcommand{\thefootnote}{\arabic{footnote}}
\setcounter{footnote}{0}

\linespread{1.1}
\parskip 4pt


\tableofcontents

\section{Introduction}

Most of the interest in the AdS/CFT correspondence \cite{Maldacena:2001pb} and its gauge/gravity generalizations
(see the books \cite{Nastase:2015wjb,Ammon:2015wua} for reviews) comes from the description it gives of nonperturbative 
quantum field theories via perturbative string theory in gravitational backgrounds. However, in the cases of most interest, 
which are closer to the real world, and to QCD in particular, the holographic map is less understood, so it is worth 
exploring ways to understand it better. One way is by using the Penrose limit, leading to the pp wave correspondence, 
originally defined in \cite{Berenstein:2002jq} for the $AdS_5\times S^5$ case, leading to a maximally supersymmetric 
type IIB pp wave \cite{Blau:2001ne,Blau:2002dy} on the string side
(see the book \cite{Nastase:2017cxp} for more details about the pp wave correspondence). 
In the Penrose limit one field theory side one 
restricts to a sector of long operators, of a large global (R-)charge, corresponding to long discretized strings in the pp wave.

Cases of more interest are confining theories like the ones of Klebanov-Strassler \cite{Klebanov:2000hb} and 
Maldacena-N\'{u}\~nez \cite{Maldacena:2000yy} (also Polchinski-Strassler \cite{Polchinski:2000uf}) in 3+1 dimensions. 
The Penrose limits of these theories were first analyzed in \cite{Gimon:2002nr} where in the IR the Klebanov-Strassler
case gave a theory of heavy hadrons, described by long gauge invariant operators and dubbed "annulons" due to the
ring structure of the resulting hadrons. The Maldacena-N\'{u}\~nez case was argued to be qualitatively similar, though 
more difficult to analyze. 

In of 2+1 dimensions, the ${\cal N}=1$ supersymmetric confining case similar to the above is the case of the 
Maldacena-Nastase (MNa) model, for NS5-branes wrapped on $S^3$ with a twist, a case that also has spontaneous 
susy breaking. The analysis of the Penrose limit and the resulting "annulon-like" long operators for hadrons was started in 
\cite{Bertoldi:2004rn,Bigazzi:2004fc}, based on the ideas in \cite{Gimon:2002nr} ,
but was not completed until our previous paper \cite{Nastase:2021qvv}. 

A very puzzling issue in holography has been the understanding of T-duality. In the case of Abelian T-duality of 
$AdS_5\times S^5$, the
interpretation has been in terms of a circular quiver field theory coming from NS5-branes and D4-branes
\cite{Witten:1997sc,Fayyazuddin:1999zu,Alishahiha:1999ds}, but the rules of the T-duality map action on the 
field theory dual were not very clear. After the inclusion of RR-charged fields in non-Abelian T-duality 
\cite{Sfetsos:2010uq}, in \cite{Lozano:2016kum,Lozano:2016wrs} and many subsequent papers, it was shown 
that the field theory corresponds to an infinite linear quiver, but again, the T-duality map in field theory was not very 
clear. On the other hand, the role of T-duality in the Penrose limit was pioneered in \cite{Alishahiha:2002ev,Mukhi:2002ck}. 
In \cite{Itsios:2017nou}, the Penrose limit method was applied in order to understand better the Abelian and non-Abelian 
T-duals of $AdS_5\times S^5$, and to see what is their effect in field theory. 

In this paper, we consider the application of the Penrose limit method on the T-dual of the MNa model, where the
T-duality is applied in one of the directions of the $S^3$ on which the 5-branes are wrapped and twisted. The goal
is to understand better the effect of this T-duality on the effective (2+1)-dimensional confining ${\cal N}=1$ susy field theory 
and its hadronic states. 

The paper is organized as follows. In section 2 we review the MNa solution, present its T-dual and analyze the resulting 
supersymmetry. In section 3 we consider the Penrose limit of the T-dual of the MNa gravity solution, and quantize strings 
in the background. In section 4 we explain the orbifold field theory dual to the MNa T-dual, and construct the "spin chain"
for the "annulon-like" hadrons in 2+1 dimensions. In section 5 we conclude.

\section{The MNa solution and its T-dual}

\subsection{MNa solution and generalization}

The Maldacena-Nastase (MNa) \cite{Maldacena:2001pb}
gravity background of 10-dimensional type IIB supergravity, dual to NS5-branes wrapped 
on an $S^3$, with a twist that preserves ${\cal N}=1$ supersymmetry, is given by (writing only the NS-NS fields: metric, 
B-field and dilaton)
\bea
ds^2_{\rm 10, string}&=&d\vec{x}_{2,1}^2+\a' N\left[d\rho^2+R^2(\rho)d\Omega_3^2+\frac{1}{4}\left(\tilde w_L^a-A^a\right)^2\right]\cr
H=dB&=&N\left[-\frac{1}{4}\frac{1}{6}\epsilon_{abc}\left(\tilde w_L^a-A^a\right)\left(\tilde w_L^b-A^b\right)\left(\tilde w_L^c-A^c\right)
+\frac{1}{4}F^a\left(\tilde w_L^a-A^a\right)\right]+h\cr
h&=&N\left[w^3(\rho)-3w(\rho)+2\right]\frac{1}{16}\frac{1}{6}\epsilon_{abc}w_L^aw_L^bw_L^c\cr
A&=&\frac{w(\rho)+1}{2}w_L^a\cr
\Phi&=&\Phi(\rho)\;.\label{MNsol}
\eea

Here $w^a_L$ and $w_R^a$ are the left- and right-invariant one-forms on the $S^3$ on which the NS5-branes 
are wrapped, respectively. Its metric is $d\Omega_3^2=w^a_Lw^a_L=w^a_Rw^a_R$. 
$\tilde w^a_L$ and $\tilde w^a_R$ are the corresponding forms on the sphere transverse to the NS5-branes, i.e., at infinity, 
$S^3_\infty$. The one-forms are parametrized by angles $\psi,\theta,\phi$ as 
\bea
w_L^{1}  =\sin\psi d\theta-\sin\theta\cos\psi d\phi\;,&
w_L^{2}  =\cos\psi d\theta+\sin\theta\sin\psi d\phi\;,&
w_L^{3}  =d\psi+\cos\theta d\phi\cr
w_R^{1}  =-\sin\phi d\theta+\sin\theta\cos\phi d\psi\;,&
w_R^{2}  =\cos\phi d\theta+\sin\theta\sin\phi d\psi\;,&
w_R^{3}  =d\phi+\cos\theta d\psi\;.\cr
&&
\eea

The functions $w(\rho),R^2(\rho)$ and $\Phi(\rho)$ are found perturbatively or numerically, subject to boundary conditions in the 
UV. Note that one can also S-dualize the solution to a solution describing D5-branes, using ($\Phi_D=-\Phi$ is the S-dual 
dilaton)
\be
ds^2_{\rm 10, D, string}=e^{\Phi_D(\rho)}ds^2_{\rm 10, string}\;,\;\; H^{(D)}=e^{\Phi_D(\rho)}H.
\ee

The generalization by Canoura et al. \cite{Canoura:2008at}, has metric 
\bea
ds_{st}^{2}&=&e^{\Phi}\left(dx_{1,2}^{2}+ds_{7}^{2}\right)\cr
ds_{7}^{2}&=&N\left[e^{2g}d\rho^{2}+\frac{e^{2h}}{4}\left(w_{L}^{i}\right)^{2}
+\frac{e^{2g}}{4}\left(\tilde{w}_{L}^{i}-\frac{1}{2}(1+w)w_{L}^{i}\right)^{2}\right]\;,
\eea
and the RR 3-form field is 
\bea
F_{3}&=&\frac{N}{4}\left\{ \left(w_{L}^{1}\wedge w_{L}^{2}\wedge w_{L}^{3}-\tilde{w}_{L}^{1}\wedge\tilde{w}_{L}^{2}
\wedge\tilde{w}_{L}^{3}\right)-\frac{\gamma^{\prime}}{2}d\rho\wedge\tilde{w}_{L}^{i}\wedge w_{L}^{i}-\right.\cr
&&\left.-\frac{(1+\gamma)}{4}\epsilon_{ijk}\left[w_{L}^{i}\wedge w_{L}^{j}\wedge\tilde{w}_{L}^{k}-\tilde{w}_{L}^{i}
\wedge\tilde{w}_{L}^{j}\wedge w_{L}^{k}\right]\right\}.
\eea

The functions $e^{2g}, e^{2h}$ (generalizing $R^2(\rho)$) and $w(\rho),\Phi(\rho)$ and $\gamma(\rho)$
are in the IR (at small $\rho$)
\bea
e^{2g}(\rho) & =&g_{0}+\frac{\left(g_{0}-1\right)\left(9g_{0}+5\right)}{12g_{0}}\rho^{2}+\ldots\cr
e^{2h}(\rho) & =& g_{0}\rho^{2}-\frac{\left.3g_{0}^{2}-4g_{0}+4\right)}{18g_{0}}\rho^{4}+\ldots\cr
w(\rho) & =&1-\frac{3g_{0}-2}{3g_{0}}\rho^{2}+\ldots\cr
\gamma(\rho) & =&1-\frac{1}{3}\rho^{2}+\ldots\cr
\Phi(\rho) & =&\Phi_{0}+\frac{7}{24g_{0}^{2}}\rho^{2}\;,
\eea
and for $g_{0}=1$, i.e, for $g=0$, we find the MNa original solution, S-dualized to the D5-brane solution.

\subsection{T-dual of (generalization of) MNa}

The T-duality of the above background was performed in \cite{Araujo:2014kda}, by use of Buscher's T-duality rules, 
\begin{equation}
\begin{gathered}
e^{2\bar{\Phi}}=\frac{e^{2\Phi}}{\left|G_{99}\right|}\quad\widetilde{G}_{99}=\frac{1}{G_{99}}\\
\widetilde{G}_{MN}=G_{MN}-\frac{G_{9M}G_{9N}-B_{9M}B_{9N}}{G_{99}}\quad\widetilde{G}_{9M}=\frac{1}{G_{99}}B_{9M}\\
\widetilde{B}_{MN}=B_{MN}-2\frac{B_{9[M}G_{N]9}}{G_{99}}\quad\widetilde{B}_{M9}=-\frac{G_{M9}}{G_{99}}\;,
\end{gathered}
\label{eq: buscher rules}
\end{equation}
and the corresponding rules for the RR form fields,
\begin{subequations}
\begin{equation}
\begin{split}
C^{(2n+1)}_{M_1\dots M_{2n+1}}&=C^{(2n+2)}_{M_1\dots M_{2n+1}\tilde{\phi}}+(2n+1)
B_{[M_1|\tilde{\phi}|}C^{(2n)}_{M_2\dots M_{2n+1}]}\\
&+2n(2n+1)B_{[M_1|\tilde{\phi}|}g_{M_2|\tilde{\phi}|}C^{(2n)}_{M_3\dots M_{2n+1}]\tilde{\phi}}/g_{\tilde{\phi}\tilde{\phi}}
\end{split}
\end{equation}
\begin{equation}
C^{(2n+1)}_{M_1\dots M_{2n}\tilde{\phi}}=C^{(2n)}_{M_1\dots M_{2n}}-2n g_{[M_1|\tilde{\phi}|}
C^{(2n)}_{M_2\dots M_{2n}]\tilde{\phi}}/g_{\tilde{\phi}\tilde{\phi}}
\;.
\end{equation}
\end{subequations}

Defining 
\bea
 \Delta&=&e^{\Phi}Ne^{2g}\cr
  \Sigma&=&\frac{e^{\Phi}}{4}N\left(e^{2h}+\frac{e^{2g}}{4}(1+w)^{2}\right)\equiv e^{\Phi}
  \widetilde{\Sigma}\equiv e^{2\tilde{\Phi}}\tilde{\Sigma}^{2}\cr
 \Omega&=&\frac{e^{\Phi}}{4}Ne^{2g}\equiv\frac{\Delta}{4}\cr
 \Xi &=&-\frac{e^{\Phi}}{8}(1+w)Ne^{2g}\equiv-\frac{\Omega}{2}(1+w)\;,
\eea
though in the following we will restrict ourselves to the strict MNa case $g_0=1$ (so in the IR $e^{2g}=1+{\cal O}(\rho^4)$, 
$e^{2h}=\rho^2+{\cal O}(\rho^4)$, $w=1-\rho^2/3+{\cal O}(\rho^4)$, $\Phi=\Phi_0+7\rho^2/24$),
corresponding to $b=1/3$ in the notation used in \cite{Nastase:2021qvv},
after the T-duality on $\phi_1$, one finds the metric 
\bea
d\tilde{s}_{st}^{2}&=&e^{2\tilde{\Phi}}\frac{N}{4}\left(e^{2h}+\frac{e^{2g}}{4}(1+w)^2\right)d\vec{x}_{1,2}^{2}
+\Delta d\rho^{2}+\frac{1}{\Sigma}d\phi_{1}^{2}+\Sigma\left(d\theta_{1}^{2}+\sin^{2}\theta_{1}d\psi_{1}^{2}\right)\cr
& & +2\Xi\left[\cos\left(\psi_{1}-\psi_{2}\right)d\theta_{1}d\theta_{2}-\sin\left(\psi_{1}-\psi_{2}\right)
 \sin\theta_{2}d\theta_{1}d\phi_{2}\right.\cr
& & -\sin\left(\psi_{1}-\psi_{2}\right)\sin\theta_{1}\cos\theta_{1}d\psi_{1}d\theta_{2}\cr
& & \left.+\left(\cos\theta_{2}\sin^{2}\theta_{1}-\cos\theta_{1}\sin\theta_{1}\sin\theta_{2}\cos\left(\psi_{1}
-\psi_{2}\right)\right)d\psi_{1}d\phi_{2}+\sin^{2}\theta_{1}d\psi_{1}d\psi_{2}\right]\cr
& & +\left(\Omega-\frac{\Xi^{2}}{\Sigma}\sin^{2}\left(\psi_{1}-\psi_{2}\right)\sin^{2}\theta_{1}\right)d\theta_{2}^{2}\cr
& & +\left(\Omega-\frac{\Xi^{2}}{\Sigma}\left[\sin\theta_{1}\sin\theta_{2}\cos\left(\psi_{1}-\psi_{2}\right)+
\cos\theta_{1}\cos\theta_{2}\right]^{2}\right)d\phi_{2}^{2}\cr
& & +\left(\Omega-\frac{\Xi^{2}}{\Sigma}\cos^{2}\theta_{1}\right)d\psi_{2}^{2}+\cr
& & +2\left(\Omega\cos\theta_{2}-\frac{\Xi^{2}}{\Sigma}\cos\theta_{1}\left[\sin\theta_{1}\sin\theta_{2}\cos\left(\psi_{1}
-\psi_{2}\right)+\cos\theta_{1}\cos\theta_{2}\right]\right)d\phi_{2}d\psi_{2}\cr
& & -2\frac{\Xi^{2}}{\Sigma}\left[\sin\theta_{1}\sin\theta_{2}\cos\left(\psi_{1}-\psi_{2}\right)+\cos\theta_{1}
\cos\theta_{2}\right]\sin\left(\psi_{1}-\psi_{2}\right)\sin\theta_{1}d\theta_{2}d\phi_{2}\cr
& & -2\frac{\Xi^{2}}{\Sigma}\sin\left(\psi_{1}-\psi_{2}\right)\sin\theta_{1}\cos\theta_{1}d\theta_{2}d\psi_{2}\;,
\eea
where $e^{2\tilde{\Phi}}=e^{2\Phi}/\Sigma$ is the T-dual dilaton, the coefficient of $dx^2_{1,2}$ equals $e^\Phi$ (the original 
dilaton), the B-field is 
\bea
\mathcal{B}&= & -\left\{ \cos\theta_{1}d\psi_{1}\wedge d\phi_{1}+\frac{\Xi}{\Sigma}\sin\left(\psi_{1}
-\psi_{2}\right)\sin\theta_{1}d\theta_{2}\wedge d\phi_{1}\right.\cr
& & \left.+\frac{\Xi}{\Sigma}\left[\sin\theta_{1}\sin\theta_{2}\cos\left(\psi_{1}-\psi_{2}\right)+
\cos\theta_{1}\cos\theta_{2}\right]d\phi_{2}\wedge d\phi_{1}+\frac{\Xi}{\Sigma}\cos\theta_{1}d\psi_{2}\wedge d\phi_{1}\right\} \;,\cr
&&\label{Bfield}
\eea
and the angles with the index ``1'' parameterize the internal (on which the branes are wrapped)
$S^3$ and the angles with index ``2'' parameterize the $S^3_\infty$. There are also R-R fields $F_2\neq 0, F_4\neq 0$ but, 
since we will not be considering fermions in the string worldsheet action, we will not write them here. 

When T-dualizing, we have the consistency condition \cite{Rocek:1991ps},
\be
\int d\phi_{1}\wedge d\tilde{\phi}_{1}=\left(2\pi\right)^{2}\;,
\ee
where we denote the original coordinate as $\phi_{1}$ and the T-dual
as $\tilde{\phi}_{1}$. 

If the $\phi_1$ direction is orbifolded by $\mathbb{Z}_{N_1}$, giving a periodicity $2\pi/N_1$ for $\phi_1$, 
the above condition implies that $\tilde{\phi}_{1}\in[0,2\pi N_{1}]$. This is the case that is relevant for us, since otherwise 
it is hard to make sense of the T-duality in the field theory dual \cite{Itsios:2017nou} 

\subsection{Supersymmetry}\label{secsusy}

To understand the amount of supersymmetry, we need to either construct the Killing spinor equations, or use symmetry 
arguments. 

If we did not have any orbifolding, we would be back in the MNa case, analyzed in \cite{Nastase:2021qvv}.
As it is, we need to understand whether, when taking the orbifold of $S^3/\mathbb{Z}_{N_1}$, where $\mathbb{Z}_{N_1}$ 
acts on an $S^1$ fiber inside $S^3$, any supersymmetry survives. In general, for a quotient manifold $X/G$, the 
unbroken susy from $X$ is the one that is invariant under $G$.

To see that, we will analyze the action of $U(1)_{S^1}$ symmetry (reduced by $\mathbb{Z}_{N_1}$) 
on the supercharges, reduced down to 2+1 dimensions.

The 5+1 dimensional theory on the NS5-branes has ${\cal N}=(1,1) $ supersymmetries. The 10-dimensional Majorana-Weyl 
spinor decomposes under the decomposition $SO(9,1)\rightarrow SO(5,1)\times SO(4)$ as 
\be
\mathbf{16}  \rightarrow  (\mathbf{4},\mathbf{1},\mathbf{2})\oplus(\bar{\mathbf{4}},\mathbf{2},\mathbf{1})\;,
\ee
where we used the fact that $SO(5,1)\simeq SU(4)$ and $SO(4)\simeq SU(2)_{A}\times SU(2)_{B}$.
Compactifying on the (internal) $S^3$ means the decomposition
\bea
SO(5,1)\times SO(4) & \rightarrow & SO(2,1)\times SU(2)_{T}\times SU(2)_{A}\times SU(2)_{B}\Rightarrow\cr
(\mathbf{4},\mathbf{1},\mathbf{2})\oplus(\bar{\mathbf{4}},\mathbf{2},\mathbf{1}) & \rightarrow & 
(\mathbf{2,\mathbf{2}},\mathbf{1},\mathbf{2})\oplus(\mathbf{2},\mathbf{2},\mathbf{2},\mathbf{1}).
\eea

Finally, twisting means considering only the diagonal subgroup, $(SU(2)_{T}\times SU(2)_{A})_{\rm diag}$ (embedding the 
gauge group in the spin connection). This gives the decomposition
\bea
SO(2,1)\times SU(2)_{T}\times SU(2)_{A}\times SU(2)_{B} & \rightarrow & 
SO(2,1)\times\left(SU(2)_{T}\times SU(2)_{A}\right)_{\rm diag}\times SU(2)_{B}\cr
\Rightarrow(\mathbf{2,\mathbf{2}},\mathbf{1},\mathbf{2})\oplus(\mathbf{2},\mathbf{2},\mathbf{2},\mathbf{1}) & \rightarrow & 
(\mathbf{2,\mathbf{2}},\mathbf{2})\oplus(\mathbf{2},\mathbf{1},\mathbf{1})\oplus(\mathbf{2},\mathbf{3},\mathbf{1})\;.
\eea

Since there is only one $SO(2,1)$ spinor that is invariant ($\mathbf{1}$) under the twisted connection (diagonal group), 
we have ${\cal N}=1$ supersymmetry.

In our case, $U(1)_{S^1}\subset SU(2)_{\rm diag}$, which means that the remaining supercharge is invariant under 
$\mathbb{Z}_{N_1}$, and survives the orbifolding. 

This is as we want, since the orbifolding we consider shouldn't interfere with the symmetries of the solution. 

\section{Strings in Penrose limit of T-dual of MNa}

\subsection{Penrose limit of T-dual of MNa}

We want to consider the Penrose limit in order to better understand the effect of T-duality on gravity dual pairs. But then, 
the most useful Penrose limit is in the T-duality direction, $\phi_1$. As in previous cases, however (for instance, 
\cite{Nastase:2021qvv} and \cite{Araujo:2017hvi,Itsios:2017nou}), that is not consistent, and we must consider also
motion in another spatial coordinate (and of course, in time $t$). Based on what we expect from the field theory, we 
want the motion to combine with one of the other two obvious isometries of the metric, $\phi_2$ and $\psi_+$, 
where 
\be
\psi_\pm\equiv \psi_1\pm \psi_2.
\ee

We choose to mix the motion on $\phi_1$ with motion on $\phi_2$. 

In order to find a null geodesic, we also impose the usual conditions, which on our metric reduce to
\be
\d^\mu g_{\phi_2\phi_2}=0\;,\;\;\; \d^\mu g_{\phi_1\phi_1}=0.
\ee

The first observation is that, for $\mu=\rho$, since at small $\rho$, the functions in the metric depend on $\rho^2$, so 
$\d^\rho$ is proportional to $\rho$ (and the multiplying functions don't have anything special), the solution would be $\rho=0$. 
However, we can easily check that for $\rho=0$ we get singularities in the metric (metric coefficients vanish for the case of the 
other angles being on their solutions). That means that we will need to keep $\rho\neq 0$, though very small (so it will be 
almost a solution to the geodesic equation, up to vanishingly small corrections, $\rho\rightarrow 0$). But that is fine 
for the Penrose limit, since $\rho$ will be transverse to the geodesic, so will scale with $1/R\rightarrow 0$. 

For the angles (that are not isometries), we get the conditions
\bea
&&\sin\psi_-\cos\psi_-\sin^2\theta_1=0\;,\;\;\;
\sin^2\psi_-\sin\theta_1\cos\theta_1=0\cr
&&[\sin\theta_1\sin\theta_2\cos\psi_-+\cos\theta_1\cos\theta_2]\sin\theta_1\sin\theta_2\sin\psi_-=0\;,\cr
&&[\sin\theta_1\sin\theta_2\cos\psi_-+\cos\theta_1\cos\theta_2](\cos\theta_1\sin\theta_2\cos\psi_--\sin\theta_1\cos\theta_2)=0\;,\cr
&&[\sin\theta_1\sin\theta_2\cos\psi_-+\cos\theta_1\cos\theta_2](\sin\theta_1\cos\theta_2\cos\psi_--\cos\theta_1\sin\theta_2)=0\;,
\eea
which we see are all satisfied by 
\be
\theta_1=\pi/2,\psi_-=\pi/2,\;,\; \theta_2={\rm arbitrary}. 
\ee
(There are other solutions, but they lead to even more singular coefficients for the metric, so we will ignore them). 

We then consider the null geodesic defined by $\rho=0,\theta_1=\pi/2, \psi_-=\pi/2$, with $\theta_2$ and $\psi_+$ arbitrary 
{\em angles} (so not interacting with the geodesic). Then they will not be rescaled by $1/R$ in the Penrose limit. 
We note then that the coefficient of $d\theta_2^2$ in the metric is singular (vanishes) at the strict geodesic point, but in the 
Penrose limit that is fine, since $\theta_2$ is not rescaled. The same comment also applies for the coefficient of $d\psi_+^2$
in the metric. Since, of course, there is no true singularity in the metric, the apparent singularity {\em must be} of the 
type of the one near the center in polar coordinates, and this is indeed what we will find. 

We define the coefficient of $d\phi_2^2$ in the metric as the function 
\be
f\equiv \Omega-\frac{\Xi^{2}}{\Sigma}\left[\sin\theta_{1}\sin\theta_{2}\cos\left(\psi_{1}-\psi_{2}\right)+
\cos\theta_{1}\cos\theta_{2}\right]^{2}.
\ee

Then, at the geodesic point we have the values
\be
f\rightarrow f_0=\Delta_0=\Sigma_0=\Omega_0=\frac{\Xi^2_0}{\Sigma_0}=-\Xi_0=\frac{e^{\Phi_0}N}{4}.
\ee

For later use, we write the full metric at the geodesic location,
\bea
ds^2&=& e^{\Phi_0}d\vec{x}_{1,2}^2+\frac{1}{f_0}d\phi_1^2+f_0\left[4d\rho^2+d\theta_1^2+0\cdot d\theta_2^2+0\cdot d\psi_+^2
\right.\cr
&&\left.+d\psi_-^2+d\phi_2^2-2\cos\theta_2 d\phi_2d\psi_- +2\sin\theta_2 d\phi_2 d\theta_1\right].\label{metricgeodesic}
\eea

The Lagrangian for a particle moving on this geodesic is (considering the values of the functions {\em at the geodesic})
\bea
L&=&-e^{2\tilde{\Phi}_0}\tilde{\Sigma}_0\frac{\dot{t}^2}{2}+\frac{\dot{\phi}_1^2}{2\Sigma_0}
+f_0\frac{\dot{\phi}_2^2}{2}\cr
&=& -e^{\Phi_0}\frac{\dot t^2}{2}+\frac{\dot \phi_1^2}{2f_0}+f_0\frac{\dot\phi_2^2}{2}\;,
\eea
and the null condition for the geodesic means $L=0$.

The cyclic coordinates are
$t, \phi_1$ and $\phi_2$ and the conserved momenta are 
\bea
\frac{\partial L}{\partial\dot{t}}=-e^{\Phi_0}\dot{t}=-E,\quad\frac{\partial L}{\partial\dot{\phi}_1}
=\frac{\dot{\phi}_1}{f_0}=-J_1\;,\;\; \frac{\d L}{\d \dot\phi_2}=f_0\dot\phi_2=J_2.\label{eq:integral of motions}
\eea

The $L=0$ (geodesic being null) constraint then gives
\be
\frac{J_2^2}{f_0}=\frac{E^2}{e^{\Phi_0}}-\Sigma_0J_1^2.\label{Jcond}
\ee

Considering $u$ as the affine parameter on the null geodesic, we make the change of coordinates
\bea
dt&=& \dot t du = \frac{E}{e^{\Phi_0}}du\cr
d\phi_1&=&\dot \phi_1 du +dw=J_1 f_0 du+dw\cr
d\phi_2&=& \dot \phi_2 du +dv=\frac{J_2}{f_0}du +dv\;,
\eea
after which the metric becomes
\bea
ds^2&=& du^2\left[-E^2e^{\Phi-2\Phi_0}+J_1^2\frac{\Sigma_0^2}{\Sigma}+\frac{f}{f_0^2}J_2^2\right]+\frac{dw^2}{\Sigma}
+fdv^2\cr
&&+e^{\Phi}d\vec{x}_{1,2}^2+\Delta d\rho^2+g_{lm}dy^l dy^m+2du\left[J_1 dw\frac{\Sigma_0}{\Sigma}
+J_2dv\frac{f}{f_0}+g_{\phi_2l}dy^l\right]\;,
\eea
where $y^l=(\theta_1,\psi_-,\theta_2,\psi_+)$. 

Now we finally make the coordinate change 
\be
f_0dV\equiv J_1 dw +\frac{f}{f_0} J_2 dv+g_{\phi_2l}dy^l\Rightarrow dv =\frac{f_0}{f J_2}\left[f_0dV-J_1 \frac{\Sigma_0}{\Sigma}dw
-g_{\phi_2l}dy^l\right]\;,
\ee
which gets rid of the mixing term with $du$, and puts the metric in a form appropriate for taking the rescaling and  
the Penrose limit,
\bea
ds^2&=& du^2\left[-E^2e^{\Phi-2\Phi_0}+J_1^2\frac{\Sigma_0^2}{\Sigma}+\frac{f}{f_0^2}J_2^2\right]+\frac{dw^2}{\Sigma}
+e^{\Phi}d\vec{x}_{1,2}^2+\Delta d\rho^2+g_{lm}dy^l dy^m\cr
&&+2du dV +\frac{f_0^2}{f J_2^2}\left[J_1^2\frac{\Sigma_0^2}{\Sigma^2}dw^2+2J_1\frac{\Sigma_0}{\Sigma}g_{\phi_2l}dw dy_l
+g_{\phi_2l}dy^l g_{\phi_2m}dy^m\right.\cr
&&\left.+\;{\rm terms}\;\;{\rm with}\;\;dV\right].
\eea

We have not written the terms with $dV$ in the brackets ($f_0^2dV^2-2f_0 dV J_1\frac{\Sigma_0}{\Sigma}dw
-2f_0 dV g_{\phi_2l}dy^l$), since they will be scaled away in the Penrose limit, as we can easily check. 
Note that in $g_{lm}dy^l dy^m$, {\em in the neighbourhood of the geodesic} we have the metric (\ref{metricgeodesic}), 
where there are no $\psi_+,\theta_2$ components, so there we have a {\em de facto} reduction to $y^l=(\psi_-,\theta_1)$ 
only. This is good, since otherwise the term with $dVg_{\phi_2\psi_+}d\psi_+$ and $dV g_{\phi_2\theta_2}d\theta_2$ 
would contribute. Then there we also have
\be
g_{\phi_2l}: g_{\phi_2\psi_-}=-\cos\theta_2 f_0\;,\;\;
g_{\phi_2\theta_1}=\sin\theta_2f_0.
\ee

Now we see that indeed the Penrose limit rescaling needs to be, as we advertised, 
\bea
&&u=U\;,\;\; V =\frac{V'}{R^2}\;,\;\; \theta_2=\theta'_2\;,\psi_+=\psi'_+\;,\cr
&& w=\frac{w'}{R}\;,\;\; \rho =\frac{\rho'}{R}\;,\;\; x_i=\frac{x_i'}{R}\;,\;\; \theta_1-\frac{\pi}{2}=\frac{\theta'_1}{R}\;,\;\;
\psi_--\frac{\pi}{2}=\frac{\psi'_-}{R}\;,\label{scaling}
\eea
and for simplicity of notation we remove the primes after the procedure. Moreover, we can, as usual, identify the overall 
scale of the metric, $f_0=e^{\Phi_0}N/4=g_sN/4$, with $R^2$. 

We need to consider the coefficients of $d\theta_2^2$ and $d\psi_+^2$ at subleading order in $\rho,\theta_1, \psi_-$, and 
the same for the coefficient of $du^2$ in the metric above, since these will all contribute to the Penrose limit. We first obtain 
\bea
\Sigma&\simeq& \frac{e^\Phi N}{4}\left(1+\frac{2\rho^2}{3}\right)\simeq f_0\left(1+\frac{23\rho^2}{24}\right)\;, 
e^\Phi\simeq e^{\Phi_0}\left(1+\frac{7\rho^2}{24}\right)\cr
\Xi&\simeq&-\frac{e^\Phi N}{4}\left(1-\frac{\rho^2}{6}\right)\simeq-f_0\left(1+\frac{\rho^2}{8}\right)\;,\cr
 \Omega
&\simeq& \frac{e^\Phi N}{4}\simeq f_0\left(1+\frac{7\rho^2}{4}\right)\cr
\Omega-\frac{\Xi^2}{\Sigma}\sin^2\psi_-\sin^2\theta_1&\simeq & \frac{e^\Phi N}{4}\left(\rho^2+\delta\psi_-^2+\delta \theta_1^2\right)
\simeq f_0\left(\rho^2+\delta\psi_-^2+\delta \theta_1^2\right)\cr
f&\simeq& \frac{e^\Phi N}{4}\left[1-(\sin\theta_2\delta \psi_-+\cos\theta_2\delta \theta_1)^2\right]\cr
&\simeq& f_0\left[1+\frac{7\rho^2}{24}-(\sin\theta_2\delta\psi_-+\cos\theta_2\delta\theta_1)^2\right].
\eea

With the proposed Penrose scaling in (\ref{scaling}), we obtain that the metric in the $\psi_1,\psi_2$ and $\theta_2$ directions 
is, at leading order,
\be
f_0(\rho^2+\delta\psi_-^2+\delta \theta_1^2)+f_0\delta\psi_-^2+f_0\frac{\rho^2}{4}\delta\psi_+^2\;,
\ee
where $\delta\psi_-=\psi_--\pi/2$, $\delta\theta_1=\theta_1-\pi/2$. 

Finally taking the Penrose rescaling and limit, and dropping the primes on the rescaled variables, 
we obtain the metric (after using (\ref{Jcond}))
\bea
R^2ds^2&=& 2f_0dU dV -f_0\left[\frac{5}{4}J_1^2\rho^2+\left(\frac{E^2}{e^{\Phi_0} f_0}-J_1^2\right)\left(\sin\theta_2 \psi_-
+\cos\theta_2 \theta_1\right)^2\right]dU^2+e^{\Phi_0}d\vec{x}_2^2\cr
&&+f_0\left[4d\rho^2+\frac{\rho^2}{4}d\psi_+^2+(\rho^2+\psi_-^2+\theta_1^2)d\theta_2^2+d\psi_-^2+d\theta_1^2\right.\cr
&&\left.+\frac{J_1^2}{J_2^2}dw^2+2\frac{J_1}{J_2}dw\frac{(\sin\theta_2 d\theta_1-\cos\theta_2d \psi_-)}{J_2}
+\frac{(\sin\theta_2 d\theta_1-\cos\theta_2 d\psi_-)^2}{J_2^2}\right]\;,\cr
&&
\eea
where note that $E^2/(e^{\Phi_0}f_0)-J_1^2=J_2^2/f_0^2\geq 0$, so the coefficient of $dU^2$ is negative definite, as it should
be, and in the last line we have kept $J_2$ as it is, to make the formulas clearer. Since $J_2$ is a derived quantity, 
we define $J_2/f_0\equiv \tilde J_2$ for convenience. Moreover, we finally identify $R^2=f_0$ and they will drop from the metric.

Making the rotation $\sin\theta_2 d \psi_-+\cos\theta_2d\theta_1\equiv d\tilde\psi_-$, $-\cos\theta_2 d\psi_-+\sin\theta_2 d\theta_1
\equiv d\tilde \theta_1$, rescaling $x_i$ by $N/4$, $\rho=\tilde \rho/4$ and $(J_1/J_2)dw\equiv d\tilde w$, we get
\bea
ds^2&=&2dU dV -\left[\frac{5}{16}J_1^2\tilde \rho^2+\left(\frac{E^2}{e^{\Phi_0 f_0}}-J_1^2\right)\tilde \psi_-^2\right]dU^2
+d\vec{x}_2^2\cr
&&+d\tilde \rho^2+\tilde \rho^2d\left(\frac{\psi_+}{4}\right)^2+\left(\frac{\tilde \rho^2}{4}+\tilde \psi_-^2+\tilde \theta_1^2\right)
d\theta_2^2+d\tilde \psi_-^2+d\tilde \theta_1^2+d\left(\tilde w +\frac{d\tilde \theta_1}{f_0 \tilde J_2}\right)^2.\cr
&&
\eea

We note that, since we are in the limit $R^2=f_0\rightarrow \infty$, the term with $d\tilde \psi_-/(f_0\tilde J_2)\rightarrow 0$ 
drops out. Then, defining $d\tilde \rho^2+\tilde \rho^2d(\psi_+/4)^2\equiv d\vec{y}_2^2$, $\tilde \psi_-\equiv z_1, \tilde \theta_1=z_2$, 
we finally get the metric
\bea
ds^2&=& 2dU dV -\left[\frac{5}{16}J_1^2\vec{y}^2+\left(\frac{E^2}{e^{\Phi_0}f_0}-J_1^2\right)z_1^2\right]dU^2\cr
&&+d\vec{x}_2^2+d\vec{z}_2^2+d\vec{y}_2^2+\left(\frac{1}{4}\vec{y}^2+\vec{z}^2\right)d\theta_2^2+d\tilde w^2.
\eea

We can define $d\theta_2\sqrt{\vec{y}^2/4+\vec{z}^2}\equiv d\eta$, with $\eta$ a Cartesian coordinate transverse
to $\vec{y}$ and $\vec{z}$, such that it has $d\eta^2$ in the metric. 

The B-field becomes then
\bea
B&= &f_0\left(J_1 dU+\frac{dw}{f_0}\right)\wedge \left\{\delta \theta_1 d\delta\psi_--\left(1-\frac{\delta\psi_-^2}{2}-\frac{\delta \theta_1
^2}{2}-\frac{5}{6}\rho^2\right)d\theta_2^2\right.\cr
&&\left.-(\sin\theta_2\delta \psi_-+\cos\theta_2\delta \theta_1)\left(\tilde J_2 dU +\frac{dV}{\tilde J_2}-\frac{J_1dw}{f_0\tilde J_2}
-\frac{-\cos\theta_2d\psi_-+\sin\theta_2d \theta_1}{\tilde J_2}\right)\right\}.\cr
&&
\eea

There is a term of order $f_0$ that under the Penrose limit becomes of order $R^2\rightarrow\infty$, but it can be removed 
by making a gauge transformation $\delta B=d \lambda$, with 
\be
\Lambda=-f_0\left(J_1 U +\frac{w}{f_0}\right)d\theta_2. 
\ee

Then under the Penrose limit there is a term of order $f_0/R=R$, but that is $\propto dU\wedge dU$, so vanishes. 
So finally, after the Penrose limit, dropping the primes on the rescaled coordinates, as before, we get
\bea
B&=& J_1 dU\wedge \left\{\theta_1d \psi_-+\left(\psi_-^2+\theta_1^2+\frac{5}{24}\tilde \rho^2\right)d\theta_2^2
+\tilde \psi_-\left[d\tilde w\left(1+\frac{\tilde J_2^2}{J_1^2}\right)+\frac{d\tilde \theta_1}{\tilde J_2}\right]\right\}\cr
&=& J_1 dU \wedge \left\{ \theta_1d\psi_-+\left(\vec{z}^2+\frac{5}{24}\vec{y}^2\right)d\theta_2
+\tilde \psi_-\left[d\tilde w \left(1+\frac{\tilde J_2^2}{J_1^2}\right)+\frac{dz_2}{\tilde J_2}\right]\right\}.
\eea

Note that, if we define $\left(\psi_-^2+\theta_1^2+\frac{5}{24}\tilde \rho^2\right)\equiv T_1^2$ and $T_1d\theta_2\equiv 
dT_2$ (since $T_1$ is a radius and $\theta_2$ an angle), we have 
\be
H=dB= J_1 dU\wedge \left\{d\theta_1\wedge d\psi_-+dT_1\wedge dT_2+d\tilde \psi_-\wedge 
\left[d\tilde w\left(1+\frac{\tilde J_2^2}{J_1^2}\right)+\frac{d\tilde \theta_1}{\tilde J_2}\right]\right\}\;,
\ee
so has constant components in some Cartesian coordinates: $H=dU\wedge h$ gives $h=h_{ij}dX^i \wedge dX^j$, 
with $h_{ij}$ constant. 

\subsection{String quantization in the pp wave background}

To quantize the string in the pp wave background, we write, as usual, the Polyakov action in the pp wave background, 
choose the conformal gauge $\sqrt{h}h^{\alpha\beta}=\eta^{\alpha\beta}$ ($h$ is the worldsheet metric) and 
light-cone gauge $x^{+}=p_{-}\tau$, where $\tau$ is the worldsheet time, obtaining
\bea
S&=& -\frac{1}{4\pi \a '}\int d\tau \int _0^{2\pi \a' p^+}d\sigma\left[\sum_{i=1}^2(\d_a x_i)^2+\sum_{i=1}^2 (\d_a y_i)^2
+\sum_{i=1}^2(\d_a z_i)^2+(\d_a\tilde w)^2\right.\cr
&&\left.+\left(\frac{\vec{y}^2}{4}+\vec{z}^2\right)\sum_{a=1}^2(\d_a\theta_2)^2-\left(\frac{5}{16}J_1^2\vec{y}^2
+\tilde J_2^2 z_1^2\right)\right]\cr
&&-\frac{J_1}{4\pi}\int d\tau\int_0^{2\pi \a' p^+}d\sigma 
\left\{ \theta_1\d_\sigma\psi_-+\left(\vec{z}^2+\frac{5}{24}\vec{y}^2\right)\d_\sigma\theta_2\right.\cr
&&\left.+z_1\left[\d_\sigma\tilde w \left(1+\frac{\tilde J_2^2}{J_1^2}\right)+\frac{\d_\sigma z_2}{\tilde J_2}\right]\right\}.\cr
&&
\eea

For the solutions of the equations of motion we choose, as usual, free wave (Fourier mode) solutions of the type
\be
\Phi_A=\Phi_{A,0}\exp\left[-i\omega\tau+ik_A\sigma\right]\;,
\ee
where $\Phi_A$ stands for all the oscillators. Moreover, also as usual, with the rescaling by $p^+$ of the above, the 
quantization of momenta around the closed string circle in $\sigma$ gives
\be
k_{A,n}=\frac{n_A}{\a'p^+}.
\ee

Using this ansatz, the equations of motion for the oscillators become algebraic, although a bit complicated and 
unyielding to solve, so we will not attempt it. 

The B field, being of the type $h_{ij}X_i dX^j$, with $h_{ij}$ constant, as noted already, will then contribute to the equations
of motion terms with $\sim ik_A= in_A/(\a'p^+)$. These will appear in skew-diagonal contributions (proportional to 
$\epsilon^{ij}$) coupling the 
various oscillators. Like in the non-T-dualized MNa case in \cite{Nastase:2021qvv}, 
then if we take all $n_A=n$ equal, we will find that the term with $n$ in $\omega_n$ will be simply {\em added} to the mass term.  

But in any case we are only interested in the mass terms, which are independent on the B-field (since the B-field is proportional 
to $k$, so to $n$). The masses are then:
\bea
&&x_i\;,\; i=1,2: \; m=0\;,\;\;\;
\tilde w\;:\;  m=0\;,\;\;\;
\eta\;:\; m=0\;,\cr
&&y_i\;,\; i=1,2: \; m=\frac{\sqrt{5}J_1}{4}\;,\;\;\;
z_i\;,\; i=1,2:\; m=\tilde J_2.
\eea

As in \cite{Itsios:2017nou}, we will see that from the dual field theory we have only the relevant case when the 
two masses are equal, when $\sqrt{5}J_1/4=\tilde J_2$, and it will not be clear how to obtain the nontrivial parameter 
$J_1$ to vary from the field theory point of view.

\section{Field theory and spin chain}

\subsection{Orbifold field theory}

On the $N$ 5-branes we start with an $U(N)$ theory. The action of the $U(1)_{S^1}$ on the $S^3\subset \mathbb{R}^4=\mathbb{C}
^2$ defined as 
\be
|z_{0}|^{2}+|z_{1}|^{2}=1\;,
\ee
is 
\be
\begin{aligned}z_{0} & \rightarrow z_{0}\\
z_{1} & \rightarrow e^{i\alpha}z_{1}.
\end{aligned}
\ee

As mentioned, $\mathbb{Z}_{N_1}$ acts by restricting the range of $U(1)$, so by $\a\rightarrow \a/N_1$.

Then, the gauge group of the orbifolded gauge theory on the wrapped 5-branes is $U(n_{1})\times\cdots\times U(n_{N_{1}})$
(see, for instance, \cite{Lawrence:1998ja,Bershadsky:1998cb}), where 
\be
\sum_{i=1}^{M}n_{i}dim(\mathbf{r}_{i})=\sum_{i=1}^{M}n_{i}=N.
\ee

In the covering space, the group is $U(N_{1}N)$. The bosonic spectrum is as follows: 
1 gauge field,${\left(A_{\mu}^{i}\right)^I}_J$, 5 real scalars,${\left(\Phi_{\beta}^{i}\right)^I}_J$
transforming in the adjoint representation of the gauge group, 1 complex
scalar (or 2 real ones), ${\left({B^{I_i}}_{J_{i+1}}\right)^i}_{i+1}$ transforming
in the bifundamental representation of two consecutive $U(n)$'s in
the gauge group, where $\mu, \nu$ are Lorentz indices, $I,J$ are fundamental group indices ($I_i$ for the $i$'th group), 
such that $(IJ)$ is in the adjoint, $i,j$ label the $U(n)$'s, 
and $\b$ refers to other internal indices.

The quiver diagram is circular, with nodes connected by the bifundamental fields. 

Then the action for the gauge fields and the bifundamental scalars is \cite{Forcella:2009jj}
\bea
S&=&\sum_{i}\frac{k_{i}}{8\pi}S_{CS_{i}}^{\mathcal{N}=1}\left(\Gamma_{i}^{\alpha}\right)\cr
&&-\int d^3x\int d^{2}\theta_{1}\left[\sum_{W_{ij}}\operatorname{Tr}\left(\left(D^{\alpha}+i\Gamma_{j}^{\alpha}\right)W_{ij}^{\dagger}
\left(D_{\alpha}-i\Gamma_{\alpha}^{i}\right)W_{ij}\right)+\mathcal{W}^{\mathcal{N}=1}\left(W_{ij},W_{ij}^{*}\right)\right]\;,\cr
&&
\eea
where 
\bea
\mathcal{W}^{\mathcal{N}=1}\left(Y_{ij},Y_{ij}^{*}\right) & =&\mathcal{W}\left(W_{ij}\right)
+\mathcal{W}\left(W_{ij}^{*}\right)+\sum_{i,k_{i}\neq0}\frac{k_{i}}{4\pi}R_{i}^{2}\cr
\frac{k_{i}}{2\pi}R_{i} & =&\sum_{j}W_{ij}W_{ij}^{\dagger}-\sum_{k}W_{ki}^{\dagger}W_{ki}.
\eea

The CS levels are as follows:

-in the original theory, the $H_3^{RR}$ field has a $k_6$ flux on $S^3$, 
leading to the same CS level, which is reduced to $k=k_6-N/2$ by integrating 
out the fermions to have just a pure CS theory. 

-in the orbifold theory, the $H_3^{RR}$ flux is distributed between the nodes of the quiver, leading at each node 
to a CS level of $k_6 n_i/N$, reduced to $k_i=k_6 n_i/N-n_i/2$ by integrating out the fermions of the node. 
Note that then we have $\sum_i k_i=k$.

\subsection{Spin chain}

In subsection \ref{secsusy} we have seen that under the twisted reduction on $S^3$, the symmetry group 
decomposes into $SO(2,1)\times\left(SU(2)_{T}\times SU(2)_{A}\right)_{\rm diag}\times SU(2)_{B}$ and, 
as we have shown in \cite{Nastase:2021qvv}, the gauge fields result in the decomposition ${\bf (3,1,1)\oplus(1,3,1)}$ and 
the scalar in ${\bf (1,2,2)}$. Moreover, the $\phi^M$ scalar modes in ${\bf (1,2,2)}$ are written, under the $SO(4)\simeq 
SU(2)_L\times SU(2)_R$ bifundamental decomposition, as 
\be
\Phi^{\a\b}=\frac{1}{\sqrt{2}}(\sigma_M)^{\a\b}\phi^M=\frac{1}{\sqrt{2}}\begin{pmatrix}i\Phi^0+\Phi^3 & \Phi^1-i\Phi^2\\
\Phi^i+i\Phi^2 & i\Phi^0-\Phi^3\end{pmatrix}=\frac{1}{\sqrt{2}}\begin{pmatrix}W& Z^*\\ Z& -W^*\end{pmatrix}.
\ee

Under the orbifold action, the 2 real bifundamental scalars ${({B^I}_J)^i}_{i+1}$ descend from the complex $Z$ scalars, 
and the 5 real scalars ${(\Phi^i_\b)^I}_J$ at a single node $i$ 
descend from the 3 $A_a$ scalars (in the ${\bf (1,3,1)}$, from the gauge field 
decomposition), and the 2 scalars in the complex $W$. 

As explained in \cite{Nastase:2021qvv}, the spin chain before orbifolding was described in terms of 
a vacuum $|0,p^+\rangle\sim \Tr [Z^J]$, in which we inserted the 8 bosonic 
oscillators $\Phi^Q=(D_\mu, A_a, W,\bar W)$, $\mu=0,1,2$; $a=1,2,3$.

But as described in \cite{Itsios:2017nou} in the case of T-duals of $AdS_5\times S^5$, one must consider the quiver orbifold 
theory, based on the works \cite{Alishahiha:2002ev,Mukhi:2002ck}, and then consider a kind of T-dual in the quiver 
($\mathbb{Z}_{N_1}$) direction. 

As explained in \cite{Mukhi:2002ck,Itsios:2017nou} then, the vacuum of the original orbifold theory, with momentum $p=1$ and 
winding $m=0$ in the $\mathbb{Z}_{N_1}$ direction, is associated with a gauge-invariant state that has "winding" around the quiver. 
But under the T-duality, $p$ and $m$ are interchanged, so we have a state of momentum $p=0$ and winding $m=1$,
created by multiplying the bifundamental scalars (only the holomorphic ones in the complex notation, $B$ and not $\bar B$),
\be
|p=1,m=0\rangle_{\rm before \; T.d.}=|p=0,m=1\rangle_{\rm after\; T.d.}={\cal O}_{N_1}
=\frac{1}{\sqrt{\cal N}}\Tr[{B^1}_2{B^2}_3...{B^i}_{i+1}...{B^{N_1}}_1].
\ee

The 8 bosonic oscillators on these states with winding are provided by the following 8 objects:

-as usual, the 3 covariant derivatives $D_\mu$, $\mu=0,1,2$.

-the 5 real scalars in the adjoint, $\Phi^i_\b$, $\b=1,...,5$, comprised of the 3 scalars from $A_a$ (${\bf (1,3,1)}$) 
and the scalars in $W,\bar W$. 

However, we will see shortly that the energies of these objects are not the same, as they will have different values of $J$.

Then the insertions of $D_\mu$ and $\Phi^i_\b$ at zero transverse momentum give the operators:
\bea
{\cal O}^a_{D,0}&=& a^{a;\dagger}_{D,0}|p=0,m=1\rangle_{\rm after\; T.d.}=\frac{1}{\sqrt{NN_1}}\frac{1}{\sqrt{\cal N}}
\sum_{i=1}^{N_1}\Tr[{B^1}_2{B^2}_3...{B^{i-1}}_i (D_a{B^i}_{i+1})...{B^{N_1}}_1]\cr
{\cal O}^\b_{\Phi,0}&=&a^{\b;\dagger}_{\Phi,0}|p=0,m=1\rangle_{\rm after\; T.d.}=\frac{1}{\sqrt{NN_1}}\frac{1}{\sqrt{\cal N}}
\sum_{i=1}^{N_1}\Tr[{B^1}_2{B^2}_3...{B^{i-1}}_i\Phi^i_\b {B^i}_{i+1}...{B^{N_1}}_1].\cr
&&
\eea

The state of momentum $p$, distributed as a sum of mode numbers $n_q$, is (for instance for $\Phi_\b^l$ insertions)
\be
{\cal O}^\b_{\Phi,p}=a^{\b;\dagger}_{\Phi,n_q}|p;m=1\rangle_{\rm after\;T.d.}=\frac{1}{\sqrt{NN_1}}\frac{1}{\sqrt{\cal N}}
\sum_{l=1}^{N_1}\Tr[{B^1}_2{B^2}_3...{B^{l-1}}_l\Phi^l_\b {B^l}_{l+1}...{B^{N_1}}_1]e^{\frac{2\pi i l n_q}{N_1}}\;,
\ee
where we have only shown one mode number (momentum) $n_q$ insertion for simplicity, and the total momentum 
$p$ (what used to be the winding before the T-duality, i.e., in the original orbifolded theory) is the sum, 
\be
p=\sum_q n_q.
\ee

To find the correspondence with the string oscillator states, we will consider the comparison with the case before T-duality. 
The first observation that we have is that $D_\mu$ splits into $D_{x_i}$ and $D_t$, as before T-duality, but now also 
$\Phi_\b$ split: $A_a$ into $(A_1,A_2)$ and $A_3$, and also $W$ and $\bar W$.

The oscillators in this case are then reshuffled, as is their charge $J$. 
Indeed, now we are interested in the gravitational symmetry charge 
\be
J=J_1+\tilde J_2=J_{\phi_1}+J_{\phi_2}.
\ee

But $J_{\phi_1}$ corresponds now not to the $J_{\phi}$ from \cite{Nastase:2021qvv}, which was a $U(1)\subset
(SU(2)'_L\times SU(2)'_R)_{\rm diag}$ of the $S^3$ (internal), but rather it is a $U(1)$ on which we T-dualize, 
thus breaking $SO(4)'=SU(2)'_L\times SU(2)'_R$ of the $S^3$ to $U(1)_{\phi_1}\times SU(2)$. This has also the 
effect of breaking the 3 gauge fields $A_a$ into $(A_1,A_2)$ rotated by $U(1)_{\phi_1}$ and $A_3$, invariant. 
Further, $J_{\phi_2}$ is the same $J_{\tilde\phi}$ in \cite{Nastase:2021qvv}, which is a $U(1)\subset SU(2)_L$ 
of the $S_\infty^3$. But, because we don't have anymore the $J_{\tilde \psi}$ of \cite{Nastase:2021qvv}, it will 
now be convenient to choose the normalization of the charges a factor of 2 larger, so $J_{\phi_1}$ 
of $A_1,A_2$ and $D_t$ is now $+1$, and $J_{\phi_2}$ of $Z,W$ is $-1/2$. The Hamiltonian is $H=\mu(\Delta-J-E_0)$, 
with $E_0=1$ as before T-duality, and with $\mu=-1$. 

Then we get the table (with respect to the fields before T-duality)
\begin{center}
    \begin{tabular}{|c|c|c|c|c|c|c|c|c|}
    \hline
    field & $Z$ & $W$ & $\bar Z$ & $\bar W$ & $A_{1,2}$ & $A_3$ & $D_t$ & $D_{x_i}$ \\
    \hline\hline
    $\Delta$ & 1/2 & 1/2 & 1/2 & 1/2 & 1 & 1 & 1 & 1 \\
    \hline
    $J_{\phi_1}$ & 0 & 0 & 0& 0 & 1 & 0 & 1 & 0\\
    \hline
    $J_{\phi_2}$ & -1/2 & -1/2 & +1/2 & +1/2 & 0 & 0 & 0 & 0 \\
    \hline
    $J$ & -1/2 & -1/2 & +1/2 & +1/2 & 1 & 0 & 1 & 0 \\
    \hline
    $\Delta-J$ & 1 & 1 & 0 & 0 & 1 & 0 & 1 & 0\\
    \hline
    $H/(-1)=\Delta-J-E_0$ & 0 & 0 & 1 & 1 & 1  & 0 & 1 & 0\\
    \hline
    oscillator & - & $\eta$ & - & $y_1$  & $z_1,z_2$ & $\tilde w$ & $y_2$ & $x_1,x_2$\\
    \hline
    \end{tabular}
\end{center}

We have listed the oscillators that correspond to the various insertions, as follows. $D_{x_i}$ obviously 
correspond to $x_i$, and $A_{1,2}$ to $z_{1,2}$ (as they come from $\tilde \theta_1,\tilde \psi_-$); $A_3$ corresponds to
$\tilde w$ (since this comes from a combination of $\phi_1,\phi_2$), and $W$ corresponds to $\eta$, which is generated by 
the angle $\theta_2$ in a transverse direction different than the one of $Z$ (which is $\phi_2$). Finally $\bar W$ to $y_1$
(which is the transverse direction $\tilde \rho$), and $D_t$ to $y_2$, which is $\psi_+$, that includes $\psi_1$, that also 
charges the $D_t$ direction, due to the twist. 

We see that the masses of the pp wave oscillators indeed match the Hamiltonian, for the case of all masses equal 
($\sqrt{5}J_1/4=\tilde J_2$). As we already commented, it is unclear, just as in the T-dual of $AdS_5\times S^5$ case 
studied in \cite{Itsios:2017nou}, why the free parameter $J_1$ is not represented in the field theory.

\section{Discussion and Conclusions}

In this paper we have constructed the T-dual of the MNa solution, and taken the Penrose limit, in order to have a 
simpler way to study the resulting field theory dual, now of orbifold type. 
We have found the spectrum of string theory oscillators, and matched 
it to the spectrum of insertions into the field theory "annulon-like" long gauge invariant operators. The effect of T-duality 
on these operators (and thus on the hadrons associated with them) was described. 

It is still not clear in general how to quantify the effect of the T-duality (in a transverse direction to the (2+1)-dimensional 
field theory, namely on the $S^3$ that the 5-branes wrap) on general states in the orbifold field theory. We have described the 
action in the pp wave limit, which in the case of the spin chain for the annulons corresponds to the "dilute gas approximation"
(or few "impurities"), but it generalizes to an action on a generic state of the (long!) spin chain. It is also still not clear why the 
T-duality action 
on general short states is a symmetry, as is the case for the string theory on the gravity dual. 
The hadrons ("annulons") are states of the confining theory in the IR, and it is not in general clear that such a theory 
should have a symmetry. Moreover, the effect (if any) on the spontaneous susy breaking needs to be understood.


\section*{Acknowledgements}


We thank Carlos N\'{u}\~nez  for useful comments. 
The work of HN is supported in part by CNPq grant 301491/2019-4 and FAPESP grant 2019/21281-4. 
HN would also like to thank the ICTP-SAIFR for their support through FAPESP grant 2021/14335-0.
The work of MRB is supported by FAPESP grant 2022/05152-2.

\bibliography{TdualPenroseMN}
\bibliographystyle{utphys}

\end{document}